\documentclass[12pt]{article}
\usepackage{amsfonts}
\usepackage{epsfig}
\usepackage{epstopdf}

\def\bbz{{\mathbb Z}}
\def\dk{{\cal B}_K}
\begin{document}
\pagestyle{plain}
\setcounter{page}{1}

\baselineskip16pt

\begin{titlepage}

\begin{flushright}
MIT-CTP-4202
\end{flushright}
\vspace{8 mm}

\begin{center}

{\Large \bf Efficient tilings of de Bruijn and Kautz graphs\\}
\vspace{3mm}

\end{center}

\vspace{7 mm}

\begin{center}

Washington Taylor$^{1}$, Jud Leonard$^2$, 
and Lawrence C.\ Stewart$^3$

\vspace{3mm}
${}^1${\small \sl Center for Theoretical Physics\\
MIT\\
Cambridge, MA 02139, USA} \\[0.15in]
$^2${\small \sl
220 Dorset Rd\\
Newton, MA 02468, USA}\\[0.15in]
$^3${\small \sl 
Serissa Research\\
7 Erwin Road\\
Wayland, MA 01778, USA}\\

\vspace{3mm}

{\small \tt wati} at {\tt mit.edu, jud.leonard} at {\tt gmail.com, stewart} at {\tt serissa.com} \\

\end{center}

\vspace{8 mm}

\begin{abstract}
Kautz and de Bruijn graphs have
a high degree of connectivity which makes them ideal candidates for
massively parallel computer network topologies.  In order to realize a
practical computer architecture based on these graphs, it is useful to
have a means of constructing a large-scale system from smaller,
simpler modules.  In this paper we consider the mathematical problem
of uniformly tiling a de Bruijn or Kautz graph.  This can be viewed as
a generalization of the graph bisection problem.  We focus on the
problem of graph tilings by a set of identical subgraphs.  Tiles
should contain a maximal number of internal edges so as to minimize
the number of edges connecting distinct tiles.  We find necessary and
sufficient conditions for the construction of tilings.  We derive a
simple lower bound on the number of edges which must leave each tile,
and construct a class of tilings whose number of edges leaving each
tile agrees asymptotically in form with the lower bound to within a
constant factor.  These tilings make possible the construction of
large-scale computing systems based on de Bruijn and Kautz graph
topologies.
\end{abstract}
\vspace*{0.2in}

\vspace{1cm}
\begin{flushleft}
\today
\end{flushleft}
\end{titlepage}
\newpage

\section{Introduction}

The family of graphs known as de Bruijn graphs
\cite{db},
Kautz graphs \cite{Kautz}, and generalized de Bruijn and Kautz graphs
\cite{ii1, ii2, rpk, dch}, have closely
related mathematical structure.  These graphs exhibit a high degree of
connectivity, which makes them natural candidates for massively
parallel computer network topologies \cite{bp}.  In particular, for
graphs of degree $K$ and diameter $N$, de Bruijn and Kautz graphs
achieve the largest possible asymptotic number of vertices $(\sim
K^N)$ \cite{pz}.

The efficient connectivity of de Bruijn and Kautz graphs makes it
difficult to partition these graphs into collections of subgraphs in
such a way as to minimize the number of edges connecting the subgraphs
to one another.  Precisely such a partition is desirable, however, in
order to physically realize a computer network topology based on a
graph of this type.  The problem of constructing a computer based on a
de Bruijn or Kautz graph from a set of isomorphic subgraphs (circuit
boards) connected together by a minimal number of edges motivates the
problem we consider in this paper of tiling de Bruijn and Kautz graphs
with isomorphic subgraph tiles having a maximal number of internal
edges.

In this paper, we define the tiling problem for de Bruijn and Kautz
graphs.  This problem can be thought of as a generalization of the
graph bisection problem applied to these graphs, where instead of
splitting the graph into two parts we wish to split the graph into a
larger number of equal-size components.  We derive a lower bound on
the number of edges which must connect the tiles of a tiling.  Based
on the criterion that a tile decomposition should be scalable to
construct graphs of arbitrary size, we give necessary and sufficient
conditions for the construction of a uniform tiling.  We construct a
family of scalable tilings which asymptotically realize the lower
bound on connecting edges to within a constant factor.  We also
provide examples of optimal tilings for small tiles.  Finally, we show
that not only de Bruijn and Kautz graphs but also their
generalizations can be tiled in a similar way using the basic
mathematical structure underlying de Bruijn graphs.

The structure of this paper is as follows.  In section 2, we fix
notation and state the fundamental problem addressed in the paper.  In
section 3, we derive a simple upper bound on the number of edges which
can be incorporated into a tile of given size, equivalent to a lower
bound on the number of connecting edges.  In section 4, we prove
necessary and sufficient conditions for the construction of a tiling
of a de Bruijn or Kautz graph, and describe a general approach for the
construction of tilings.  Section 5 describes the generalization of
the results of section 4 to generalized de Bruijn and Kautz graphs.
Section 6 contains some examples of tilings constructed using the
method of section 4.  In section 7 we give an explicit construction of
a family of tiles which asymptotically realize the lower bound of
section 3 to within a constant factor, as the tile size becomes large.
Section 8 contains a brief discussion of the application of the tiling
approach developed in this paper to supercomputer architectures,
including a mathematical argument for the optimality of degree 3
Kautz/de Bruijn graphs in certain contexts.  Section 9 contains some
brief concluding remarks.

A related formulation of this problem was given in \cite{Viterbi},
along with some explicit decompositions of degree 2 de Bruijn graphs
into isomorphic subgraph tiles similar to those described in this
paper.  We focus in this paper on the directed graph (digraph) form of
de Bruijn and Kautz graphs.  Similar considerations could be used for
tiling the undirected forms of the graphs.

\section{Definitions and problem statement}

We begin by defining some notation.

First we define a (directed) de Bruijn graph.  The de Bruijn graph 
${\cal B}_K^N$ of degree
$K$ and diameter $N$ 
is a digraph with $K^N$ vertices.  The vertices can be labeled
by strings of $N$ integers base $K$
\begin{equation}
c_1 c_2 \cdots c_N \in \bbz_K^N \,,
\end{equation}
so $c_i \in\{0, 1, \ldots, K -1\}$.  The  de Bruijn graph 
${\cal B}_K^N$
has edges
\begin{equation}
(c_1 c_2 \cdots c_N,
c_2 c_3 \cdots c_N c_{N + 1})
\label{eq:dbe}
\end{equation}
for all values of $c_i, i \in\{1, \ldots N + 1\}$.
There is thus a natural one-to-one map between edges of ${\cal B}_K^N$
and vertices of ${\cal B}_K^{N + 1}$.  This leads naturally to an
inductive construction of ${\cal B}_K^{N + 1}$ as a family of iterated
line digraphs.

A (directed)  Kautz graph ${\cal K}_K^{N + 1}$ of degree $K$ and diameter $N + 1$
can be defined similarly as a digraph with $(K +1)K^{N}$
vertices labeled by strings
\begin{equation}
s_0  s_1 \cdots  s_N \in \bbz_{K + 1}^{N + 1} \,, \; s_i \neq s_{i + 1} \,,
\end{equation}
where the integers $s_i$
are taken base $K + 1$, but subject to the condition that
adjacent integers must
be distinct.  The edges of ${\cal K}_K^{N + 1}$ are given in a similar
fashion to (\ref{eq:dbe}) by
\begin{equation}
\{(s_0 s_1 \cdots s_N,
s_1 s_2 \cdots s_N s_{N + 1})| \; s_i \in \bbz_{K +1} \; s_i \neq s_{i
  + 1} \} \,.
\end{equation}

\begin{figure}
\begin{center}
\begin{picture}(200,100)(- 100,- 50)
\put(10,10){\makebox(0,0){\epsfig{file=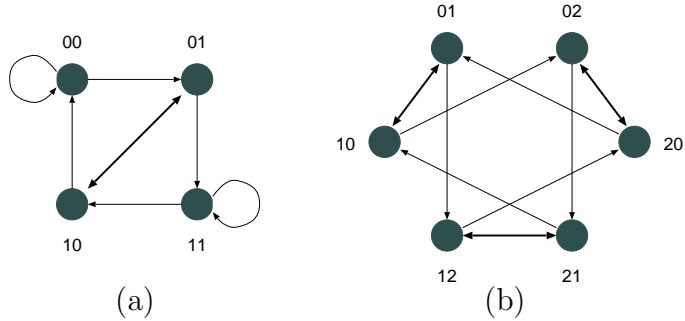,width=9cm}}}
\put(-70,-50){\makebox(0,0){(a)}}
\put(70,-50){\makebox(0,0){(b)}}
\end{picture}
\end{center}
\caption[x]{\footnotesize  The de Bruijn graph ${\cal B}_2^2$ (a)
and Kautz graph  ${\cal K}_2^2$ (b) of degree $2$ and diameter 2.}
\label{f:dk22}
\end{figure}

\begin{figure}
\begin{center}
\begin{picture}(200,100)(- 100,- 45)
\put(5,15){\makebox(0,0){\epsfig{file=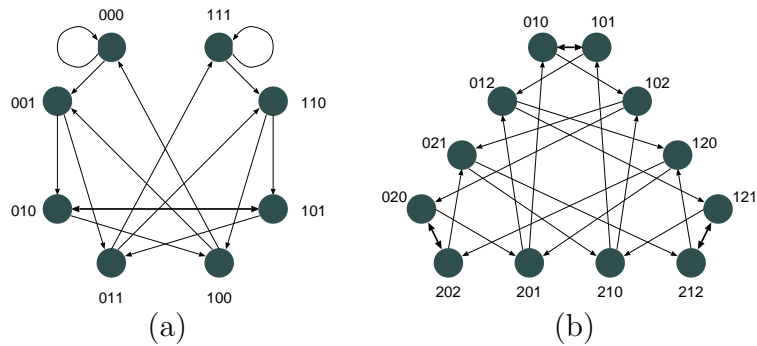,width=10cm}}}
\put(-77,-50){\makebox(0,0){(a)}}
\put(77,-50){\makebox(0,0){(b)}}
\end{picture}
\end{center}
\caption[x]{\footnotesize  The de Bruijn graph ${\cal B}_2^3$ (a)
and Kautz graph  ${\cal K}_2^3$ (b) of degree $2$ and diameter 3.}
\label{f:dk23}
\end{figure}

\begin{figure}
\begin{center}
\begin{picture}(200,140)(- 100,- 47)
\put(5,17){\makebox(0,0){\epsfig{file=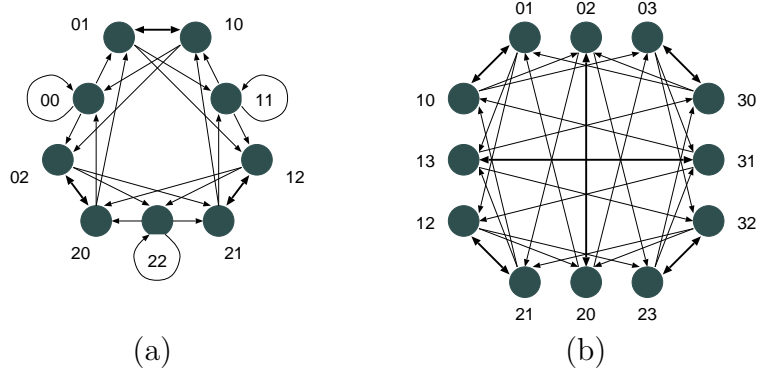,width=10cm}}}
\put(- 82,-55){\makebox(0,0){(a)}}
\put(82,-55){\makebox(0,0){(b)}}
\end{picture}
\end{center}
\caption[x]{\footnotesize  The de Bruijn graph ${\cal B}_3^2$ (a)
and Kautz graph  ${\cal K}_3^2$ (b) of degree $3$ and diameter 2.}
\label{f:dk32}
\end{figure}

Some simple examples of de Bruijn and Kautz graphs for small values of
the degree $K$ and diameter are depicted in
Figures~\ref{f:dk22}-\ref{f:dk32}

Kautz graphs and de Bruijn graphs both have uniform in-degree and
out-degree $K$.
Kautz graphs and de Bruijn graphs are closely related.  Indeed, if we
take a Kautz graph ${\cal K}_K^{N + 1}$ with vertices $s_0 \cdots
s_N$, we can map the vertices to the vertices of the de Bruijn graph
${\cal B}_K^N$ through
\begin{equation}
r:s_0 \cdots s_N \rightarrow c_1 \cdots c_N, \;
c_i = s_i-s_{i -1} -1 ({\rm mod}\;K + 1).
\label{eq:kd-homomorphism}
\end{equation}
This map is a (many to one) graph homomorphism, since if
$(s_0 \cdots s_N, s_1 \cdots s_{N +1})$ is an edge of
${\cal K}_K^{N +1}$ it follows immediately that the associated
$(c_1 \cdots c_N, c_2 \cdots c_{N +1})$ is an edge of
${\cal B}_K^{N}$.  We can thus label every vertex in the Kautz graph
by $s_0v$, where $s_0 \in\bbz_{K + 1}$ and
$v =c_1 c_2 \cdots c_N$ is a vertex in a de Bruijn graph ${\cal
  B}_K^N$.  The edges of the Kautz graph in this notation are
\begin{equation}
(s_0 c_1 c_2 \cdots c_N,
[s_0 + c_1 + 1] c_2c_3 \cdots c_{N + 1})
\label{eq:relabeling}
\end{equation}
where $[s_0 + c_1 + 1]$ is taken modulo $K + 1$.  
We will use this relationship between
Kautz and de Bruijn graphs in the following section.
\vspace*{0.1in}

\noindent
{\bf Notation}: if a directed graph has an edge $(u, v)$ we say that $v$ is
a {\it child} of $u$ and $u$ is a {\it parent} of $v$.  We denote the
set of children of $u$ by $C (u)$ and the set of parents of $v$ by $P (v)$.
\vspace*{0.1in}

We are interested in finding tilings of de Bruijn and Kautz graphs with certain
properties, where a tiling is defined as follows:
\vspace*{0.05in}

\noindent
{\bf Definition}: A {\it tiling} of a graph $G$ is a one-to-one (on
vertices) graph embedding $t: I \times V_T \rightarrow V_G$ taking the
product of an index set $I =\{1, 2, \ldots,  | V_G |/| V_T |\}$ and the vertices $V_T$ of a smaller graph
$T$ (the {\it tile}) into the vertices of the graph $G$ such that if
$(x, y)$ is an edge of $T$ then $(t (i, x), t (i, y))$ is an edge of
$G$, $\forall i$.
\vspace*{0.1in}

We wish to find tilings of the de Bruijn and Kautz graphs ${\cal
B}_K^N$ and ${\cal K}_K^{N + 1}$ such that the number of edges of
the tile $T$ is maximized.  This problem is motivated from the problem
of designing a computer with a de Bruijn or Kautz topology, where we
wish to construct the machine using a set of identical boards (the
tiles), minimizing the number of wires connecting the boards by
maximizing the number of on-board connections corresponding to edges
of $T$.
\vspace*{0.1in}
\noindent
{\bf Problem 1} For fixed $K, N$, and tile size $S = | V_T |$
such that $S | K^N$,
find a  tiling of ${\cal B}_K^N$ which maximizes $|E_T|$, the number
of edges in $T$.
\vspace*{0.1in}

This problem has previously been formulated in the case $K = 2$, and
some example tilings described in \cite{Viterbi}.

\vspace*{0.1in}
\noindent
{\bf Problem 2} For fixed $K, N$, and tile size $S | (K + 1) K^N$,
find a tiling of ${\cal K}_K^{N + 1}$ which maximizes $|E_T|$.
\vspace*{0.1in}

\vspace*{0.1in}
\noindent
{\bf Definition}: a tiling is {\it scalable} if the associated tile
$T$ can be used to tile a de Bruijn or Kautz graph of any size $K^N
$or $(K + 1)K^N \geq
|V_T|$ (for  fixed $K$).
\vspace*{0.1in}

This property is useful practically because machines of different
sizes can be built with the same modular component.
The condition that the tiling be scalable also enables us to make more
general statements about the set of allowed tilings.
\vspace*{0.1in}

\noindent
{\bf Problem 3}: For a fixed degree $K$ and tile size $S$,
find a tile $T$ with $S = | V_T |$ and
maximum $|E_T|$
which leads to a scalable tiling
for either de Bruijn or Kautz graphs.
\vspace*{0.1in}

When $K$ is prime, any tile must have size
$K^M$ for some $M$.  For non-prime $K$, it may be possible to find
tiles of other sizes giving scalable tilings.  We will focus in this
paper on tiles with sizes which are powers of $K$.  Such tiles can be
constructed for all $K$.

\vspace*{0.1in}
\noindent
{\bf Definition:} We define the maximum number of edges $E_T$ which
can be realized in any tile which has $K^M$ vertices and 
which gives a scalable tiling for de Bruijn and Kautz graphs of degree
$K$ to be
$\hat{E}_{K, M}$.

\vspace*{0.1in}
\noindent
{\bf Definition:} We define the minimum possible number of ``broken'' edges in
any tile with $K^M$ vertices to be
$\hat{B}_{K, M}= K^{M + 1}-\hat{E}_{K, M}$.
\vspace*{0.1in}

In this paper we construct a class of tilings which turn out to have
the following useful property 

\vspace*{0.1in}
\noindent
{\bf Definition}: a tiling has the {\it parallel routing} property if for
each vertex $x \in V_T$, the children of $t (i, x)$  
are mapped
under $t^{-1}$ to vertices in $I \times V_T$ with the same set of $V_T$
values  (for each $i$).
\vspace*{0.1in}
In other words, defining the projection $\Pi$ of a vertex in $G$ to
the tile $T$ through
\begin{equation}
 \Pi (t (i, x)) = x\; \; \forall i
\end{equation}
the parallel routing property states that
\begin{equation}
\Pi (C (t (i, x))) = \Pi (C (t (i', x)))
\label{eq:parallel}
\end{equation}
for any $i, i'$ in the index set $I$.
\vspace*{0.1in}

This property is useful practically because it simplifies the wiring
of a machine based on a tiling with this property.  

\section{Upper bound on internal tile edges}
\label{sec:asymptotic}

In this section we derive an upper bound on the number of edges in the
tiles of a tiling, equivalent to a lower bound on the number of edges
which must leave each tile.  We assume here that tiles are of size $K^M$; a
similar bound can be found for tiles with sizes not of this form for
composite $K$ by the same analysis.  The lower bound found here takes a simple
asymptotic form as $M \rightarrow \infty$ for fixed $K$.  We focus on
de Bruijn tilings, but analysis of Kautz tilings gives the same form
for the bound.

\vspace*{0.1in}
\noindent
{\bf Theorem 1}: The number of edges achievable in a tile which
leads to scalable tilings has an upper bound
\begin{equation}
\hat{E}_{K, M} \leq K^{M + 1} -
{\rm max}_{N}  \frac{1}{N} \left( K^{M + 1} -K^{2M + 1-N} \right)
\end{equation}

\vspace*{0.1in}
\noindent
{\bf Proof}:

We find this upper bound by considering a set of paths which leave the
tile; we bound the number of these paths which cross any given edge,
which gives us a lower bound on the number of edges which must leave
the tile.  This basic method, described in \cite{Leighton}, is used in
\cite{rttv} to construct a lower bound on the edge bisection width of
de Bruijn and Kautz graphs.

Consider a tiling of the de Bruijn graph ${\cal B}_K^N$ by tiles of
size $K^M$.  For a given tile $T$, consider the $K^M (K^N -K^M)$ paths
of length $N$ which go from a node in $T$ to a node in the de Bruijn
graph outside tile $T$.  Each of these must cross at least one edge
which begins at a node in $T$ and ends at a node outside $T$.  We
refer to such at edge as an edge ``leaving'' the tile $T$.  Each edge
leaving the tile is traversed by at most $N K^{N -1}$ of the paths
under consideration (since the edge fixes $N + 1$ of the $2 N$ $c_i$'s
defining the path, but may appear at any of $N$ places in the path).
Thus, the number of edges leaving the tile ($K^{N + 1}-E_T$) is at
least
\begin{equation}
K^{N + 1}-E_T\geq \frac{K^M(K^N -K^M)}{N K^{N -1}} 
= \frac{1}{N} \left( K^{M + 1} -K^{2M + 1-N} \right)
\label{eq:e-bound}
\end{equation}
for any choice of paths, and for any tile $T$.
Thus we have shown
\begin{equation}
\hat{B}_{K, M} \geq \frac{1}{N} \left( K^{M + 1} -K^{2M + 1-N} \right)
\label{eq:b-bound}
\end{equation}
for any $N$,
which proves the theorem.
\vspace*{0.1in}

\noindent
{\bf Corollary}
For fixed $K$ and large $M$ the lower bound on
the number of edges leaving the tile goes asymptotically as
\begin{equation}
\hat{B}_{K, M} \geq
\frac{1}{M} \left( K^{M + 1} \right)
(1 - {\cal O} (\log M/M)) \,.
\label{eq:lower-bound}
\end{equation}

\vspace*{0.1in}
\noindent
{\bf Proof}:

By taking $N = M +\lfloor \log_K M \rfloor$, we have
\begin{equation}
\hat{B}_{K, M} \geq
\frac{1}{N} \left( K^{M + 1} -K^{2M + 1-N} \right)
\sim
\frac{1}{M} \left( K^{M + 1} \right)
(1 - {\cal O} (\log M/M)) \,.
\end{equation}
In Section \ref{sec:efficient} we explicitly construct a class of
tilings which realize this asymptotic bound to within a factor of 2.

The asymptotic lower bound we have found here for the number of broken
edges is similar in form to that found in \cite{rttv} for  bisection of
de Bruijn and Kautz graphs.  Those authors found that the edge
bisection width $b_e$ satisfies the asymptotic inequalities
\begin{equation}
\frac{K^{N + 1}}{2 N}  
(1-{\cal O} (1/K^{2 N})) \leq b_e (K, N)
\leq
\frac{2 K^{N + 1}}{N}  
(1+{\cal O} (N/K^{ \sqrt{N}/2})) \,.
 \label{eq:bisection-bounds}
\end{equation}
Note that the asymptotic form of the lower bounds agrees for $K = 2, N
= M + 1$, where the tiling and bisection problems are equivalent.

\section{Structure  of  tiles}

In this section we prove a necessary condition which a tile of size
$K^M$ must satisfy to lead to a scalable tiling.  We then prove a
related but slightly stronger sufficient condition for scalable
tilings, by giving a general construction of a class of tilings based
on any tile which satisfies the sufficient condition.

We begin with a few definitions

\vspace*{0.1in}
\noindent
{\bf Definition}: A  {\it stratification} of a directed
graph $G$ is a map
$\sigma:V_G \rightarrow {\bf Z}$ from the vertices of $G$ into the
integers such that for every edge $(u, v)$ in $G$, we have
\begin{equation}
\sigma (u) = \sigma (v) + 1 \,.
\label{eq:stratification-rule}
\end{equation}

We refer to $\sigma (u)$ as the {\it level} of the vertex $u$ in the
stratified graph $G$.  By convention, we take the smallest level in a
stratification to be 0 by shifting all levels equally.
We refer to a graph which admits a stratification as {\it stratifiable}.
\vspace*{0.1in}

\noindent
{\bf Definition}: A {\it loop} is a directed graph
which is topologically equivalent to a circle.
\vspace*{0.1in}

Note that the edges in a loop need not have compatible orientations.
It is straightforward to show that a loop is stratifiable if and only
if it contains an equal number of edges of each orientation.
For example, the loop defined by the graph containing 3 vertices $u,
v, w$ and edges $\{(u, v), (v, w), (u, w)\}$ is unstratifiable since a
stratification would give
$\sigma (u) = \sigma (v) + 1 = \sigma (w) + 2$ from the first two
edges and $\sigma (u) = \sigma (w) + 1$ from the third edge.

\vspace*{0.1in}
\noindent
{\bf  Lemma 1}: A graph $G$ is stratifiable if and only if it contains
no unstratifiable loops as subgraphs.

\vspace*{0.1in}
\noindent
{\bf Proof}: If $G$ contains an unstratifiable loop $L$ then $G$ is
clearly not itself stratifiable since any stratification of $G$ would
provide a stratification of $L$.  If $G$ contains no unstratifiable
loops, then we can explicitly construct a stratification of $G$ by
choosing a single reference vertex in each connected component of $G$
and assigning the level of each other vertex $w$ in $G$ by
implementing the rule (\ref{eq:stratification-rule}) on each of the
edges needed to reach $w$ on any path from a reference vertex,
allowing the path to traverse edges in either direction.  In the
absence of unstratifiable loops, this assignment is path independent
and provides a stratification of $G$.
\vspace*{0.1in}

We can now prove the following
\vspace*{0.1in}

\noindent
{\bf Theorem 2}: A necessary condition for a tile $T$ of size $K^{M}$
to give scalable tilings of ${\cal B}_K^N$ and ${\cal K}_K^{N + 1}$
for all $N \geq M$ is that $T$ must be stratifiable.

\vspace*{0.1in} \noindent {\bf Proof}:
We begin by assuming that $T$ is unstratifiable and therefore
contains an unstratifiable loop.  We will show that this assumption
leads to the conclusion that $T$ cannot be used to construct
scalable tilings, thus proving the theorem by contradiction.

Assuming that a given tile $T$ contains an unstratifiable loop $L$ of
length $n$, we write $L$ in terms of the sequence of vertices in $T$
it traverses.
\begin{eqnarray}
L & = & [u_1, u_2, \ldots, u_{n}, u_{n + 1} =u_1] \label{eq:loop-l}\\
& &u_{i + 1} \in C (u_i), i \in l_+   \nonumber\\
& &u_{i + 1} \in P (u_i), i \in l_- \nonumber
\end{eqnarray}
where $l_+, l_-$ denote the sets of indices of vertices associated
with edges $(u_i, u_{i + 1})$
which are traversed in a forward/backward direction in the loop
$L$, so that $l_- \cap l_+ =\{\}, l_- \cup l_+ =\{1, 2, \ldots, n\}$.
Since $L$ is unstratifiable, we must have
\begin{equation}
| l_+ | \neq | l_-| \,.
\end{equation}

Now, imagine that the tile $T$  gives a scalable
tiling of de Bruijn graphs ${\cal B}_K^N$ for $N \geq M$ but nonetheless
contains the unstratifiable loop $L$ given in  (\ref{eq:loop-l}).  In a
tiling of the graph  ${\cal B}_K^N$, there will be $K^{N -M}$ nodes
$v$ which project to each node $u \in V_T$ through $u = \Pi (v)$.
In particular, there will be $K^{N -M}$ nodes $v^{(i)} =t (i, u_1)$ with
$\Pi (v^{(i)}) = u_1$.  Choose any such node $v_1 = c_1 \cdots c_N$ which
can be written as $t
(i_0, u_1) = v_1$ for some fixed
$i_0$.  Then we can define a loop in ${\cal B}_K^N$ through
\begin{equation}
v_i = t (i_0, u_i)
\end{equation}
whose vertices must satisfy
\begin{eqnarray}
v_{i + 1} & \in & C (v_i), \; \;{\rm if} \;i\in l_+, \\
v_{i + 1} & \in & P (v_i), \; \;{\rm if} \; i\in l_-
\end{eqnarray}
Since $| l_+ | \neq | l_- |$, when we follow the loop around, the string
of characters defining $v_1$ has been shifted to the left $| l_+ | -|
l_-|\neq 0$ times.  But we have returned to $v_1$ after going around the
full loop; this implies that the bulk of the word $v_1 = c_1 \cdots
c_N$ is invariant under shifting left by $| l_+ | -| l_- |$.
Specifically,
\begin{equation}
c_i = c_{i + \Delta}, \;{\rm where} \; \Delta= | l_+ | -| l_- |, \; \;i
> n,  \; \;i
\leq N - n -\Delta
\end{equation}
But only $K^{2 n + \Delta}$ nodes have this property (such nodes are
completely defined by giving $c_1 \cdots c_n, c_{N -n-\Delta+1} \cdots
c_N$), so we cannot have 
$K^{N -M}$ nodes with $\Pi (v) = u_1$ when $N > M+ 2n + \Delta$.  Thus, the
assumption of an unstratifiable loop is incompatible with the
assumption of a scalable tiling for $T$, so we conclude that if
$T$ can be used to construct a scalable tiling it cannot contain an
unstratifiable loop, and therefore must be stratifiable.  The theorem
is thus proven.
\vspace*{0.2in}

The condition of stratifiability is not by itself sufficient to
guarantee that a tile gives rise to a scalable tiling.   In some
cases, stratifiable loops can form obstructions to the construction of
a tiling.  We now will proceed to prove that scalable tilings can be
constructed on stratifiable tiles with certain types of loops.  First,
however, we need a few more definitions and a lemma.

\vspace*{0.1in}

\noindent
{\bf Lemma 2}: A tile of size $S = K^M$ giving scalable tilings of
${\cal
B}_K^N$ and ${\cal K}_K^{N + 1}$  for $N \geq M$ must be a
stratifiable subgraph of
${\cal B}_K^M$
\vspace*{0.1in}

\noindent
{\bf Proof}: The tile must be stratifiable by Theorem 2.  It must be a
subgraph of ${\cal
B}_K^M$ since it must tile ${\cal
B}_K^N$ with $N = M$.
Note that since $S = K^M$, the subgraph contains all vertices of
${\cal B}_K^M$, but will only contain a subset of the edges.
\vspace*{0.1in}

We can thus associate with any scalable tiling of a de Bruijn or Kautz graph
$G ={\cal
B}_K^N$ or ${\cal K}_K^{N + 1}$  a map
$\Pi:V_G \rightarrow V_{{\cal B}_K^M}$.  For the tilings which we will
construct, this map extends to a complete graph homomorphism
$\Pi:G \rightarrow {\cal B}_K^M$.  
We define several properties for
maps of this type
\vspace*{0.1in}

\noindent
{\bf Definition}:  A map $\Pi:G \rightarrow {\cal B}_K^M$ has  the
{\it parent distribution} property when
\begin{equation}
\Pi (P (u)) = P (\Pi (u)), \; \; \forall u \in G
\label{eq:pdp}
\end{equation}
\vspace*{0.1in}

\noindent
{\bf Definition}:  A map $\Pi:G \rightarrow {\cal B}_K^M$ has  the
{\it child distribution} property when
\begin{equation}
\Pi (C (u)) =  C (\Pi (u)), \; \; \forall u \in G
\label{eq:cdp}
\end{equation}

Note in particular that if $\Pi:G \rightarrow {\cal B}_K^M$ has both
the parent and child distribution properties, then for any vertex $u
\in G$ with $\Pi(u) =d_1
\cdots d_M \in {\cal B}_K^M$, there is a unique parent $p_{u, x}\in P
(u)$ of $u$ with $\Pi (p_{u, x}) = xd_1 \cdots d_{M -1}$ and a unique child
$c_{u, x}\in  C
(u)$ of $u$ with $\Pi (c_{u, x}) = d_2 \cdots d_{M} x$ for each $x \in
{\bf Z}_K$.
\vspace*{0.1in}

We now prove by example that there exist maps 
$\Pi:G \rightarrow {\cal B}_K^M$ with
$G ={\cal
B}_K^N$ or ${\cal K}_K^{N + 1}$ which have both the parent and child
distribution properties, for any $N \geq M$

For a de Bruijn graph ${\cal B}_K^N$, we begin by defining the $k$th
{\it discrete 
differentials}.  For each vertex $v = c_1 \cdots c_N$ we define
\begin{eqnarray}
d^0_i (v)  &= & c_i \nonumber\\
d^k_{i} (v) & = & c_{i + k} - c_i \; ({\rm mod}\; K),  \; \;
i = 1, \ldots, N -k, \; k > 0 \,. \label{eq:dcc}
\end{eqnarray}
Let us fix $N\geq M$  and take $k = N -M$.
Then the discrete 
differentials define a map
\begin{equation}
d: c_1 \cdots c_N \rightarrow d_1 \cdots d_{N -k}, \;\;\;\;\;
d_i = d^k_i (c_1 \cdots c_N)
\end{equation}
which is a graph homomorphism
\begin{equation}
d:{\cal B}_K^N \rightarrow{\cal B}_{K}^{M}
\label{eq:dd-map}
\end{equation}
since if $(c_1 \cdots c_N, c_2 \cdots c_{N + 1})$ is an edge of ${\cal
  B}_K^N$ then $(d_1 \cdots d_{N-k}, d_2 \cdots d_{N-k + 1})$ is an
  edge of ${\cal B}_K^{N -k}$.
It is straightforward to check that this graph homomorphism has both
  the child and parent distribution properties.   
Note that using this homomorphism we can relabel the vertices of the graph
 $G ={\cal B}_K^N$
  using the first $k$ $c$'s and the $d$'s
\begin{equation}
V_{G} =\{c_1 \cdots c_{k} d_1 \cdots d_{N -k}\} \,.
\label{eq:labeling-d}
\end{equation}

Combining the discrete differentials on the de Bruijn graph
${\cal B}_K^N$ with the homomorphism
(\ref{eq:kd-homomorphism}) 
gives us a map 
\begin{equation}
\tilde{d}:{\cal K}_K^{N + 1} \rightarrow{\cal B}_K^M
\label{eq:dd-Kautz}
\end{equation}
defined through $\tilde{d}(u) = d (r (u))$, with $r$ defined as in
(\ref{eq:kd-homomorphism}).  
Again, it is easy to
verify that this graph homomorphism has the child and parent
distribution properties.
Using this homomorphism, we
can label the nodes in a Kautz graph
$H ={\cal K}_K^{N + 1}$ through
\begin{equation}
V_{H} =\{s_0 c_1 \cdots c_{k} d_1 \cdots d_{N -k}
| s_0 \in\bbz_{K + 1}, c_1, \ldots, d_{N -k} \in\bbz_K\} \,.
\label{eq:labeling-k}
\end{equation}
\vspace*{0.1in}

\noindent
{\bf Definition}: We define the {\it height} of a stratifiable loop
$L$ to be the difference $\sigma_{\rm max} -\sigma_{{\rm min}}$
between the largest and smallest levels in a loop.
\vspace*{0.1in}

We can now prove the following
\vspace*{0.1in}

\noindent
{\bf Theorem 3} Given a stratified tile $T$ of size $K^{M}$ which is a
subgraph of ${\cal B}_K^{M}$ (containing all vertices but not all
edges
of  ${\cal B}_K^{M}$), tilings of $G ={\cal B}_K^N$ and ${\cal
K}_K^{N + 1}$ can be constructed for all $N \geq M$,
as long as $T$ contains no loops of height $> M$.
Such a tiling can be constructed explicitly from any map
$\Pi:G \rightarrow \dk^M$ with the child and parent distribution properties.

\vspace*{0.1in}
\noindent
{\bf Proof}: We showed above by explicit construction that a map
$\Pi:G \rightarrow \dk^M$ with the child and parent distribution
properties exists for any $M$, for $G$ a de Bruijn or Kautz graph of
degree $K$ with $N \geq M$.  Let us take any particular such $\Pi$.
From $\Pi$ we can define the tiling constructively as follows: choose
a vertex $x_0 =d_1 \cdots d_{M}$ of the tile $T$.  Associated with
this vertex of $T$ there is a set of vertices in our de Bruijn or
Kautz graph $G$ which map to $x_0$ under $\Pi$.  We can arbitrarily
associate this set of vertices with the elements $i$ in the index set
$I$ to define $t (i, x_0)$ for all $i$.  For example, using the
discrete differential map described above, given a specific $x_0 = d_1
\cdots d_M$ we can define the index set by the leading indices before
the $d's$ on this set of vertices in the notation of
(\ref{eq:labeling-d}, \ref{eq:labeling-k}) ({\it i.e.}, $c_1 \cdots
c_k$ for de Bruijn, $s_0c_1 \cdots c_k$ for Kautz).  

We have now defined $t (i, x_0)$ for all $i$ and for a specific $x_0$
in $V_T$.
Now, if
there are any edges in $T$ containing $x_0$ ({\it i.e.} directed edges
$(x_0, x')$ or $(x', x_0)$ beginning or
ending on $x_0$), we can extend the
definition.  Say $T$ contains the edge $(x_0, x_1)$.  For each $i \in
I$, using the child distribution property, there is a unique vertex
$v_i$ in $G$ which is a child of $t (i, x_0)$ and which has $\Pi (v_i)
= x_1$.  We can thus define $t (i, x_1) = v_i$ for each $i$.  These
vertices $v_i, v_j$ are distinct for $i \neq j$, by the parent
distribution property---if $v_i = v_j$ then $t (i, x_0)$ and
$t(j,x_0)$ would both be parents of the same vertex with the same
value of $\Pi$, which is impossible given the parent distribution
property.  Instead of extending along an edge where $x_0$ is the
parent, we could equivalently extend along an edge where $x_0$ is the
child, using an equivalent argument where the roles of parents and
children are exchanged.  We continue in this way for further edges
containing either $x_0$ or $x_1$.  Each time we include a new edge we
define the map $t$ for a new value of $x$.  If the graph $T$ is
disconnected, we run out of edges before the map $t$ is completely
constructed.  In this case we choose a new vertex $x'_0$ where $t$ is
not yet defined and proceed as above with $x_0'$ in place of $x_0$.
In this fashion, we can construct the complete tiling.

It remains to be shown that the construction of the tiling we have
just presented is well-defined in that any loops existing in the tile
do not lead to incompatible definitions for the tiling by following
different paths to extend the definition of the tiling.
This is straightforward to demonstrate for any loop whose height is
less or equal to $M$. Given a loop $L \subset T$ of height $h \leq M$
we need to show that
the loop can be consistently lifted to a loop in the graph $G$ by
following the edges and using the child and parent distribution
properties as in the above construction.  Let us start at a vertex
$x_1 \in L$ such that $\sigma (y) \geq \sigma (x_1) \;\forall y \in L$.
We lift $x_1 = d_1 \cdots d_M$ to a vertex $u_1\in G$ with
$\Pi (u_1) = x_1$.  Labeling the vertices of the loop in order as
$[x_1, \ldots, x_s, x_{s + 1} = x_1]$ where $s$ is the length of the
loop we can use the 
child and parent distribution properties to lift each $x_i$ to a
vertex $u_i \in G$ as above, following the edges of the loop one after
another.  We need to show that when we return to $x_1$ the final value
of $u_{s + 1}$ is the same as $u_1$.  Assume that $G =\dk^N$---a
similar argument proceeds for Kautz graphs.  We can write
$u_1 = c_1 \cdots c_N$.  From the form of the de Bruijn graph edges,
and since $x_1$ has the lowest level of any vertex in the loop $L$, we
see that all vertices $u_i$ contain the coordinates $c_1 \cdots c_{N
  -h}$, shifted to the right by $\sigma (x_i)-\sigma (x_0)$ places.
But then, $u_{s + 1}= c_1 \cdots c_{N -h} \tilde{c}_{N -h + 1} \cdots
\tilde{c}_{N}$.  Since by the child distribution property
there is a unique $M$th child of $z_1 \cdots z_M c_1 \cdots c_{N -M}$
with $\Pi = x_1$, we must have $u_{s + 1} = u_1$ when $h \leq M$, so
the loop of height $h \leq M$ does not obstruct the construction of
the tiling.  This completes the proof of the theorem.
\vspace*{0.1in}

\noindent
{\bf Corollary}: The tilings constructed by the method of Theorem 3
have the parallel routing property.
\vspace*{0.1in}

\noindent
{\bf Proof}: This follows from the use of $\Pi$ with parent and child
distribution properties, since
\begin{equation}
\Pi (C (t (i, x))) = C (\Pi (t (i, x))) = C (x)
\end{equation}
is independent of $i$ so (\ref{eq:parallel}) holds for all $i, i', x$.
The parallel routing property can also be seen to follow immediately
from the fact that $\Pi$ is a graph homomorphism.
\vspace*{0.2in}

We have now provided results which make it possible to systematically
search for optimal solutions to Problem 3.  We know that for tile size
$S = K^M$, scalable tilings are given by stratifiable subgraphs of
$\dk^M$.  We can thus consider the set of stratifiable subgraphs of
$\dk^M$, ordered by their number of edges.  We know that all
stratifiable subgraphs with no loops of height $h > M$ can give
scalable tilings.  Thus, if a stratifiable subgraph of $\dk^M$ can be
realized with the maximal number of edges and $h \leq M$, this
provides a solution to Problem 3.  In some cases, the stratifiable
subgraph of $\dk^M$ with the maximal number of edges may have a loop
of height $h > M$.  In such a case, a further check is needed to
verify that such a subgraph gives a scalable tiling; generally this
will not be possible, but it may be possible in special cases.  In any
case, finding a subgraph of $\dk^M$ with the maximal number of edges
subject to the condition that there are no loops of height $h > M$
will give rise to a good set of scalable tilings.

We conclude this section with some further comments.

First, we note that the discrete differential construction above is
not the only way to realize a map $\Pi:G \rightarrow {\cal B}_K^M$ 
with the parent and child distribution properties.  Indeed,
consider any function 
\begin{equation}
f: {\bf Z}_K \times {\bf Z}_K \rightarrow {\bf Z}_K
\label{eq:f}
\end{equation}
with the following properties:
\begin{itemize}
\item  for each $c \in {\bf Z}_K,$ $f (c, \cdot)$ gives a one-to-one
  map from ${\bf Z}_K \rightarrow{\bf Z}_K$ ({\it i.e.}, for fixed
  $c$, $f (c, c')$ takes different values for each $c'$)
\item  for each $c' \in {\bf Z}_K,$ $f ( \cdot, c')$ gives a one-to-one
  map from ${\bf Z}_K \rightarrow{\bf Z}_K$ ({\it i.e.}, for fixed
  $c'$, $f (c, c')$ takes different values for each $c$)
\end{itemize}

If we replace $d^k_i = c_{i+ k}-c_i ({\rm mod}\ K)$ in (\ref{eq:dcc})
with any other function $d^k_i =f (c_{i + k}, c_i)$ with these
properties, we see that the associated map $\Pi:G \rightarrow {\cal
B}_K^M$ still has the child and parent distribution properties when
$G$ is either a de Bruijn or Kautz graph.  Using such a more general
map in the construction of Theorem 3 provides a more general class of
constructions of scalable tilings with the parallel routing property.

Note however that not every graph homomorphism of the form $\Pi:G
\rightarrow {\cal B}_K^M$ has the parent and child distribution
properties.  For example, while the map ${\cal B}_K^N \rightarrow{\cal
B}_K^j$ defined by taking a contiguous subset $c_ac_{a +1} \cdots c_{a +
j -1}$ of  characters in the string $c_1 \cdots c_N$
is a homomorphism, it does not have these properties, and indeed
cannot be used to define a tiling.  To see this clearly in a specific
simple case, consider the case $K = 3, N = 3$.  We cannot tile the
27-node de Bruijn graph ${\cal B}_3^3$ using tiles with vertices
addressed by $c_2 c_3$.  For example, the 3 parents of the vertex
$u=001$ have coordinates $c_0c_1 c_2 = 000, 100, 200$.  These vertices
all have the same value of $c_2 c_3 = 00$, so the graph homomorphism
taking $c_0c_1 c_2 \rightarrow c_1 c_2$ does not have the parent
distribution property.  Assume  now for example that our tile includes the
edge $(00, 01)$.  The vertex $101$, which should be associated with
the vertex $01$ on the tile has no parent with $c_2c_3 = 00$, so the
edge $(00, 01)$ cannot be contained in the tile.  Similar problems
arise with any other edge, so we cannot tile ${\cal B}_3^3$ with tiles
having vertices addressed by $c_2c_3$ and more than 0 edges.

\section{Tilings of generalized de Bruijn and Kautz graphs}
\label{sec:generalized}

In this section we describe tilings of generalized de Bruijn and
Kautz graphs.
So far, we have discussed only  de Bruijn  and Kautz graphs
of sizes $K^N$ and $(K + 1)K^N$ respectively.
Generalizations of the de Bruijn and Kautz graphs to other sizes are
described in \cite{ii1, ii2, rpk, dch}.
We show here that the construction of the previous section gives
tilings by tiles of size $S =K^M$  for generalized de Bruijn and Kautz
graphs with $V =nS$ vertices for any integer $n$.  

\subsection{Generalized de Bruijn graphs}

A generalized de Bruijn graph of degree $K$ having $V$ vertices can be
defined by taking the directed graph on vertices $0 \leq i < V$ with
edges
\begin{equation}
(i, (Ki + m) \; {\rm mod} \; V), \;\;\;\;\; \forall \; m, 0 \leq m < K \,.
\label{eq:gdb}
\end{equation}
If $V = K^N$, it is easy to see that this definition agrees with the
one in (\ref{eq:dbe}) by simply taking $c_1 \cdots c_N$ to be the base
$K$ representation of the integer $i$ for each vertex $i$.
Multiplication by $K$ simply shifts the base $K$ representation left
by one digit, adding $m$ shifts in an arbitrary new digit $c_{N +1}$
and modding by $V =K^N$ truncates to $N$ digits base $K$.
More generally, we can write any $V = nK^M$  in the form
\begin{equation}
V = F K^N,
\end{equation}
where $M \geq N$ and $F$ and $K$ are relatively prime, {\it i.e.} $(F, K) = 1$.  We
can then represent any vertex $i < V$ in mixed-base form as
\begin{equation}
f c_1 \cdots c_N, \;\;\;\;\; c_i < K, f < F\,.
\label{eq:db-mixed}
\end{equation}
As before, the transformation rule associated with edges (\ref{eq:gdb})
shifts the $c$'s left.  This transformation rule
takes $fc_1 \cdots c_N \rightarrow f' c_2 \cdots c_{N + 1}$ where
\begin{equation}
f' = (Kf + c_1) \; {\rm mod} \; F \,.
\label{eq:fmap}
\end{equation}
Since $K$ and $F$ are relatively prime, multiplication by $K$ is
invertable mod $F$, so that
for any fixed $c_1$  the map defined in (\ref{eq:fmap}) from $f
\rightarrow f'$ is a one-to-one map.  In particular, given $c_1, f'$
there exists a unique $f < F$ such that (\ref{eq:fmap}) is satisfied.
This demonstrates that the map from the generalized de Bruijn graph
with $V$ vertices to the regular de Bruijn with $K^N$ vertices given
by dropping the first digit $f$ in (\ref{eq:db-mixed}) has the parent
distribution property (\ref{eq:pdp}).  This map also clearly has the
child distribution property (\ref{eq:cdp}), which follows from the
fact that the children of any node with mixed-base representation
$fc_1 \cdots c_N$ are just $f' c_2 \cdots c_N m$ with $f'$ given by
(\ref{eq:fmap}) and $0 \leq m < K$.  Since this map has both the
parent and child distribution properties, so does any  map to
a smaller de Bruijn formed by the  composition of this map with a map
$\Pi':{\cal B}_K^N \rightarrow{\cal B}_K^M$ also having the parent
and child distribution properties.  This shows that the tiles which
can be used to construct tilings of standard de Bruijn and Kautz
graphs in Theorem 3 of the previous section can also be used to tile
generalized de Bruijn graphs.  We give
some examples of such tilings in Section \ref{sec:examples}

\subsection{Generalized Kautz graphs}

Similar to the generalized de Bruijn graphs defined through
(\ref{eq:gdb}), a Kautz graph of degree $K$ having $V$ vertices can be
defined by taking the directed graph on vertices $0 \leq i < V$ with
edges
\begin{equation}
(i, (-1 - Ki - m) \; {\rm mod} \; K), \;\;\;\;\; \forall \;m,0 \leq m < V \,.
\label{eq:gk}
\end{equation}
To relate this to the standard Kautz graph in the case $V = (K +
1)K^N$, we again introduce a mixed mode representation for each vertex
$i$, writing $V = FK^N$ with $(K, F) = 1$.  We denote the digits in
this mixed mode representation by
\begin{equation}
f \bar{c}_1 c_2 \bar{c}_3c_4 \cdots \begin{array}{c}
c_N \\ \bar{c}_N
\end{array}, \;\;\;\;\; c_i, \bar{c}_j
 < K, f < F
\label{eq:gk-mixed}
\end{equation}
where the last digit is $\bar{c}_N$ if $N$ is odd, and $c_N$ if $N$ is
even.
The barred digits are defined to be $\bar{c}_i = K -1 -c_i$, giving a
shift register representation of the vertices.  From (\ref{eq:gk}) we
see that the graph edges are given by
\begin{equation}
(f \bar{c}_1 c_2 \cdots \begin{array}{c}
c_N \\ \bar{c}_N
\end{array}, f' \bar{c}_2  {c}_3 \cdots \begin{array}{c}
\bar{c}_N \\ {c}_N
\end{array} m)
\end{equation}
where
\begin{eqnarray}
f'  & = &  (-1 - Kf - \bar{c}_1) \; {\rm mod} \; F \\
 & = &  (- Kf + c_1 -K) \; {\rm mod} \; F \,.
\end{eqnarray}
In particular, if we choose $F = K + 1$, we have $K \equiv -1 ({\rm
 mod} \; F)$, so $f' =  (f + 1 + c_1) {\rm mod} \;F$, which is precisely
the transformation rule on edges in (\ref{eq:relabeling}) if we
 identify $f = s_0$.  Thus, when $F = K + 1$ we have the standard
 Kautz graph, while for other $F$ we have a generalization.

As in the previous subsection, the invertibility of multiplication by
$K$ mod $F$ guarantees that the map to a $K^N$ node de Bruijn given by
dropping the first digit $f$ in the mixed mode representation
(\ref{eq:gk-mixed}) has the parent and child distribution properties,
and therefore the generalized Kautz graphs can also be tiled using the
methods of the previous section.  Examples are given in the following
section.

\section{Examples}
\label{sec:examples}

In this section we give some explicit examples of tilings constructed
using the discrete differential method of Theorem 3.

\begin{figure}
\begin{center}
\epsfig{file=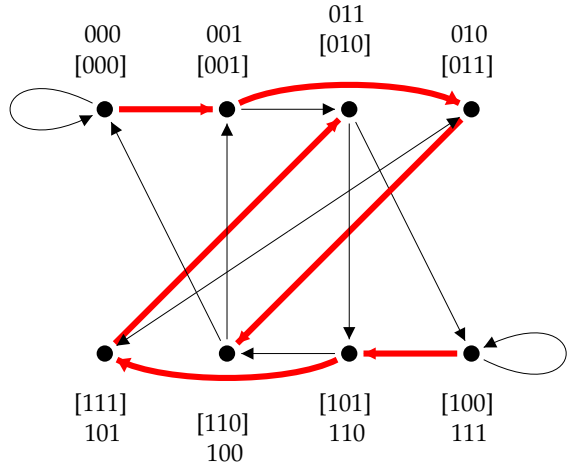,width=13cm}
\end{center}
\caption[x]{\footnotesize Optimal scalable
tiling of the de Bruijn graph
${\cal B}_2^3$ by tiles of size 4}
\label{f:d23}
\end{figure}

Let us first consider tilings of
de Bruijn graphs with $K = 2$ by tiles of size 4.  The de Bruijn graph
${\cal B}_2^2$ is shown in  Figure~\ref{f:dk22}.  Up to graph
isomorphisms there are  4 kinds of loops in this graph: 
\begin{equation}
[00 \rightarrow 00], \;\;\;
[00 \rightarrow 01 \rightarrow 10 \rightarrow 00], \;\;\;
[00 \rightarrow 01 \leftarrow 10 \rightarrow 00], \;\;\;
[00 \rightarrow 01 \rightarrow 11 \rightarrow 10 \rightarrow 00],
\end{equation}
where the second and third
loops differ only in the orientation of the link connecting
$01$ and 10.
All these loops are non-stratifiable.  Thus, there are no stratifiable
tiles given by subgraphs of ${\cal B}_2^2$ with more  edges than the
tile with 3 edges:
\begin{equation}
E_T =\{(00, 01), (01, 11), (11, 10)\}.
\label{eq:tile-22}
\end{equation}
This tile therefore gives a scalable tiling which solves Problem 3 for
$K = 2, S = 4 = 2^2$.  As an example of a tiling using this tile,
consider the tiling of ${\cal B}_2^3$ using the map $\Pi$ described
using discrete differentials (\ref{eq:dd-map}).  This tiling is
depicted in Figure~\ref{f:d23}.  In this figure the numbers in
brackets are the addresses $c_1 d_1 d_2$ for each node as in
(\ref{eq:labeling-d}).  The colored/bold links are those realized on
a tile $T$ described by (\ref{eq:tile-22}).

\begin{figure}
\begin{center}
\epsfig{file=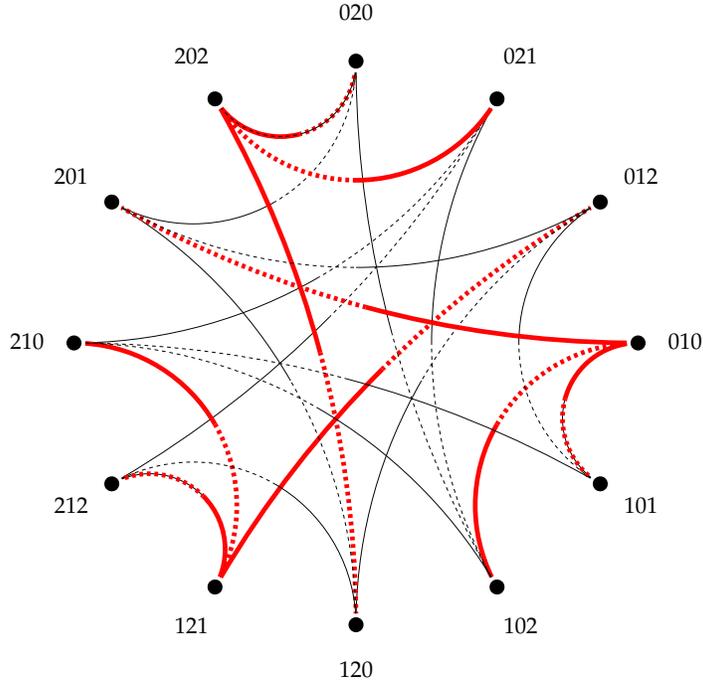,width=12cm}
\end{center}
\caption[x]{\footnotesize Optimal scalable tiling of the Kautz graph
  ${\cal K}_2^3$ by 3 tiles of size 4.  Direction of link goes from
  dotted end to solid end.}
\label{f:k23-4}
\end{figure}

Let us now consider simple examples of a Kautz graph tiling with $K =
2$.  The optimal size 4 ($M = 2$) tile for $K = 2$ again has no loops
and has only 3 edges.  We can, for example choose the stratified
subgraph of ${\cal B}_2^2$ containing edges $E_T =\{(00, 01), (10,
01), (01, 11)\}$, with $\sigma (00) = \sigma (10) = 2, \sigma (01) =
1, \sigma (11) = 0$.  We can use the discrete differential map
(\ref{eq:dd-Kautz}) $\tilde{d}:{\cal K}_2^3 \rightarrow{\cal B}_2^2$
to label the vertices of the Kautz graph ${\cal K}_2^3$ by $s_0
d_1 d_2$.  The tiling associated with this map is depicted in
Figure~\ref{f:k23-4}, where nodes are labeled by $s_0 s_1 s_2$ (the
nodes are, however, ordered according to the index $i$ from
(\ref{eq:gk})).  In this figure, again colored/bold links are realized
on a tile $T$.  The direction of each link is indicated by using
dotted lines on the outgoing part of the link and solid lines on the
incoming part of the link.

For $K = 2$, up to relabeling there is only one function of the form
(\ref{eq:f}) with the desired properties, which is the one used in the
definition (\ref{eq:dcc}) of discrete differentials.
For $K = 3$,   it is  easy to check that up to relabeling
of the integers, which is a symmetry of both the de Bruijn and Kautz
graphs, there are precisely two distinct functions with the desired properties:
\begin{equation}
f_1 (c, c') = c-c' \; ({\rm mod} \; K)
\end{equation}
and
\begin{equation}
f_2 (c, c') = c+c' \; ({\rm mod} \; K)
\end{equation}
Either of these functions can be used to construct tilings with $K = 3$.

\begin{figure}
\begin{center}
\epsfig{file=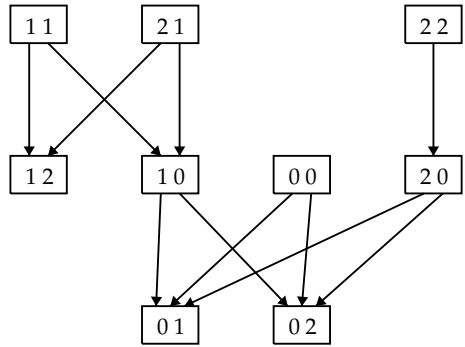,width=12cm}
\end{center}
\caption[x]{\footnotesize Optimal tile for $K = 3$, $M = 2$.  Tile has
11 internal edges  (shown) and 7 external edges (not shown)}
\label{f:tile-32}
\end{figure}

\begin{figure}
\begin{center}
\epsfig{file=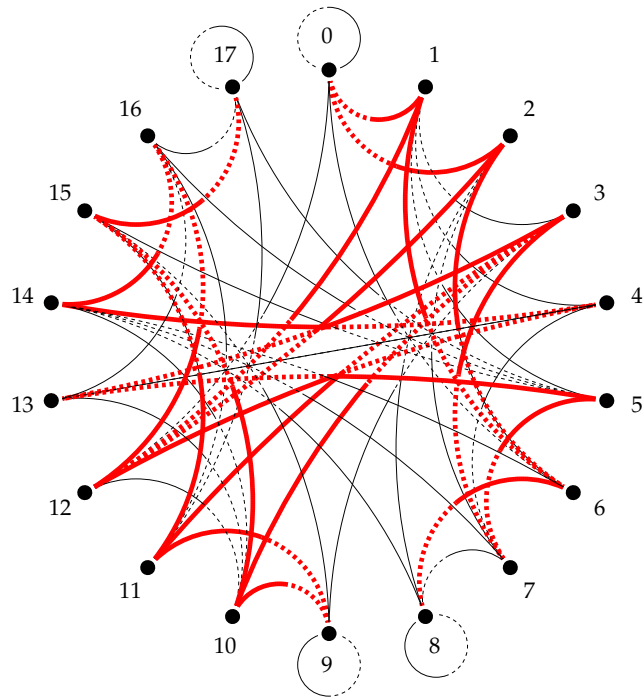,width=12cm}
\end{center}
\caption[x]{\footnotesize Optimal scalable tiling of the generalized
  de Bruijn graph with $K = 3, V = 18$ by tiles of size 9}
\label{f:d9-18}
\end{figure}

\begin{figure}
\begin{center}
\epsfig{file=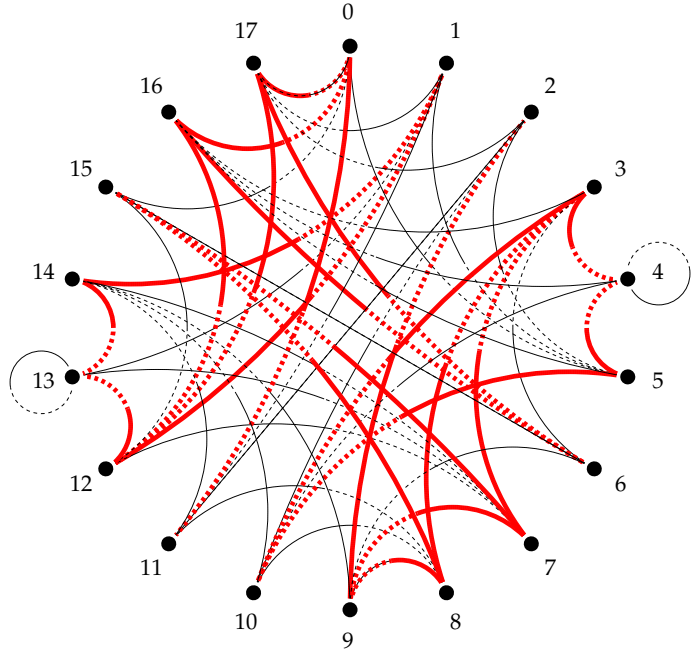,width=12cm}
\end{center}
\caption[x]{\footnotesize Optimal scalable tiling of the generalized
Kautz graph with $K = 3, V = 18$ by tiles of size 9}
\label{f:k9-18}
\end{figure}

Now, for $K = 3$, consider the tile with $M = 1$.  The de Bruijn graph
here is just the 3 vertices $0, 1, 2$ with edges going from each
vertex to each other vertex.  The largest subgraph without directed
loops has two edges, such as $0 \rightarrow 1 \rightarrow 2$.  This
defines the best tile for a scalable tiling.  Now consider $M = 2$.  A
systematic analysis of possible stratifiable subgraphs of ${\cal
  B}_3^2$ shows that the maximum number of edges compatible with
stratifiability is 11, which can be realized for example by the tile
depicted in Figure~\ref{f:tile-32}.  This tile has 4 loops of length 4
(of which 3 are homotopically independent), all of which are of height
1.
Tilings of the generalized de Bruijn and Kautz graphs with 18 nodes by
two copies of this tile are shown in Figures~\ref{f:d9-18}
and~\ref{f:k9-18}.  In  these figures the node numbers are the
numbers $i$ used in the description of generalized de Bruijn and Kautz
graphs in Section \ref{sec:generalized}.

We have carried out a systematic analysis of all tiles for small
values of $K, N$.  We have computed the optimal tiles by performing a
complete search over all possibilities in all cases where $K^N \leq 16$.
Our results are
tabulated here.

\begin{center}
\begin{tabular}{| c | c | c | | c |}
\hline
 $K$ & $N$ & $K^N$  & max $| E_T |$\\
\hline
2 & 2 & 4 & 3\\
2 & 3 & 8 & 8\\
3 & 2 & 9 & 11\\
2 & 4 & 16 & 19\\
4 & 2 & 16 & 27\\
\hline
\end{tabular}
\end{center}

Using a simple greedy-first algorithm we have found some good, but not
necessarily optimal tiles for larger values of $K^N$; as examples, for
$K = N = 3$ there is a tile with 44 internal edges, and with $K = 3, N
= 4$ there is a tile with  150 internal edges.  In the next section we
discuss tilings for arbitrarily large values of $M$

\section{Asymptotically efficient tilings}
\label{sec:efficient}

In this section we provide an explicit construction of an efficient
class of tilings.  This class of tilings realizes the asymptotic bound
(\ref{eq:lower-bound}) on edges leaving the tile to within a factor of
2.  As we discuss below, we believe that this class of tilings may in
fact be asymptotically optimal, and that the theoretical bound may be
improvable by a factor of 2.  We cannot, however, demonstrate this
conclusively at this time.  Note that for the graph bisection problem
there is a similar gap (in that case by a factor of 4) between upper
and lower bounds on edge bisection bandwidth
(\ref{eq:bisection-bounds}) for general $K, N$ \cite{rttv}.

\subsection{Local score-based tiles}
\label{sec:score-tiles}

We will now define some tiles explicitly by choosing a stratification
of the nodes in a de Bruijn graph and applying Theorem 3.
The basic idea we will use to construct efficient tilings is the
identification of local patterns in the $dd$ (discrete differential)
address space to choose the levels in the stratification of the nodes
of the tile of size $K^M$.

For each node of the de Bruijn graph
${\cal B}_K^M$ labeled by $d_1 \cdots d_M$, we
assign to each position in the node address a score between 0 and 1
according to how well the node address in the vicinity of $d_i$
matches some desired pattern.  We will use the convention that a
smaller score indicates a closer match to the desired pattern.  The
idea is that if this score is very low at some particular position
$i$ for a node $x=d_1 \cdots d_M$, then the score will generally be
low for position $i -1$ in the node addresses of all the children of
$x$.  Thus, if we choose the level of the node $x$ to be $\sigma (x)
=i$
where $i$ is the position of lowest score, then generically children
$y$ of $x$ will have lowest score at position $i-1$, and will be
assigned level $\sigma (y) = i -1$, so that the edge $(x, y)$ can
be included in the tile.

There are many ways in which we can define such a local pattern-based
score.  To give a concrete example we define such a score in a way
which leads to efficient tilings.

\vspace*{0.1in}

\noindent
{\bf Definition:} We define a {\it $K$-ary expansion
score} by interpreting the sequence
of digits beginning with $d_{i}$ as a base $K$ expansion of a  real number
between 0 and 1 in the following way
\begin{equation}
\phi_i(d_1 \cdots d_M) =
\frac{K-1-d_{i}}{K}  + \sum_{j = 1}^{\infty}  \frac{d_{i + j}}{K^{j
    + 1}}, \;\;\;\;\; \forall i: 0 \leq i \leq M
\label{eq:phi-score}
\end{equation}
where we define $d_0 = 0$  and $d_n = K -1$ for
$n > M$.  Thus for example, the node $x =0010100$ for $K = 2, M =  7$
would have $\phi_3 (x)= 0.00100\bar{1}_2 = 0.00101_2 = 5/32$.  The idea
is that the position in the node address describing the smallest
real value will determine the level of the node.
The leading digit is complemented base $K$ so that the starting point of this
lowest value is clearly marked; without this complementation the
resulting tile would have multiple broken edges whenever  a very small
number (long sequence of 0's) initiates the node address sequence $d_1
\cdots d_M$.  The boundary conditions are similarly chosen to minimize
broken edges.

With this score function, we then define a tile by defining a
stratification of nodes through
\begin{equation}
\sigma (x) = i:  \; \;
\phi_i (x) \leq \phi_j (x) \; \forall j \neq i, 0 \leq
j \leq M.
\label{eq:score-level}
\end{equation}
There are (limited)
circumstances in which two positions $i$
can have the same value of $\phi_i (x)$ for a fixed $x$, leading to
``ties'' where multiple values of $i$ satisfy (\ref{eq:score-level}).
Such ties only occur for nodes in which the last $n$ digits are all
equal to $K -1$ and all previous digits are $< K -1$.  In this case we define
\begin{equation}
\sigma ((d_1 < K -1) \cdots (d_{M-n} < K -1)
(d_{M-n + 1} = K -1) \cdots  (d_M = K -1)) =  M.
\label{eq:ties}
\end{equation}
With this assignment of levels, each of these nodes has $K -1$
children at level $M -1$ (those with $d_M\neq K -1$) and one child at
level  $M$
giving a broken edge.

\vspace*{0.1in}

\noindent
{\bf Example}: In the tile with $K = 2, M = 6$, consider the node $d_1
\cdots d_6 = 010010$.  The smallest score is found at the position
where the smallest number appears base 2, with the first bit in the
sequence beginning at that point complemented.  This is position $i =
2$, with $\phi_2 (010010)= 0.00010\bar{1}_2 = 0.00011_2 = 3/32$.  So
this node is placed at level $\sigma (010010) = 2$.  Similarly, each
of the child nodes 100100 and 100101 have the smallest $\phi_i$ at $i
= 1$ so $\sigma (100100) = \sigma (100101) = 1$, so edges (010010,
100100) and (010010, 100101) are both included in the tile.  Note that
for the node 100100, $\phi_1 (100100) = 5/64$, while $\phi_4 (100100)
= 1/8$.

We can use this method based on the $K$-ary expansion
score to construct the full tile for any $K, M$.   For each node in
the tile the level of the stratification is defined through
(\ref{eq:score-level}), with (\ref{eq:ties}) used to break ties.
\vspace*{0.1in}

\begin{figure}
\begin{center}
\epsfig{file=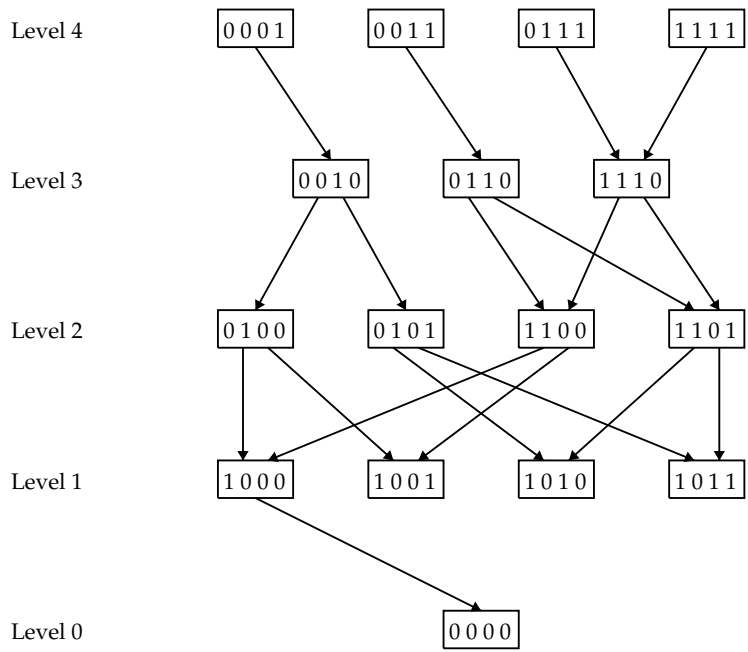,width=12cm}
\end{center}
\caption[x]{\footnotesize A tile with $K = 2, M = 4$ based on the
2-ary expansion score stratification.  Nodes with the smallest string
beginning at position $i$ (preceded by a 1) are placed on level $5-i$,
with boundary conditions and tie-break conditions as discussed in
text.
Tile is optimal, having 19 internal edges (shown) and 13 external edges (not shown).}
\label{f:tile-24}
\end{figure}

\noindent
{\bf Example}: For the case of $K = 2, M = 4,$ the $K$-ary expansion
score stratification gives the tile shown in Figure~\ref{f:tile-24}.
This tile has 19 internal edges and 13 broken edges.  Thus, this
algorithm for tile construction yields an optimal tile in this case.
\vspace*{0.1in}

It is straightforward to automate the construction of tiles according
to this system.  The numbers of internal (broken) edges for tiles with
small values of $K, M$ are listed in the following table.

\begin{center}
\begin{tabular}{|| c  || c | c | c | c |c  ||}
\hline
\hline
$K\backslash M$ & 2 & 3 & 4 & 5 & 6\\
\hline
\hline
2 & $3^*$ (5) & $8^*$ (8) & $19^*$ (13) & 42 (22) & $90$ (38)\\
3 & $10$ (17) & $41$ (40) & $146$ (97) & 485 (244) & 1559 (628)\\
4 & 23 (41) & 129 (127) & 615 (409) & 2729 (1367) & 11697 (4687)\\
5 & 44 (81) & 314 (311) & 1876 (1249) & 10414 (5211) & 55794 (22331)\\
\hline
\hline
\end{tabular}
\end{center}

In this table we denote with an asterisk (*) cases where the tile is
known to be optimal.  The $2^5$ tile with 42 internal edges may also
be optimal; we have not found a better tile using heuristic search
methods.  For most of the other small tiles, we have found slightly
better tiles than those constructed using this method with brute force
or greedy algorithm searches, or by slight modifications of the
score-based stratification algorithm described above; for example as
mentioned above for the $3^4$ tile we have found a tile with 150
internal edges.  For larger tiles, we do not know of any general
method for generating tiles with significantly more internal edges.  As we now
discuss, we expect that the asymptotic behavior of these tiles may be
the optimal achievable.  Note, however, that for larger values of $K$
the approach to the asymptotic form is slower and the tiles produced
by this method are somewhat sub-optimal.  In particular, when $K$ is
large and $M$ is small, many nodes have no digits $d_i = K -1$, which
leads to extra broken edges.  This might be improved by some heuristic
method combining such nodes into sub-tiles which are then attached to
increase total edge numbers.  But, as $M \rightarrow \infty$, for any
fixed $K$ the fraction of such nodes goes to 0 exponentially as $[(K
  -1)/K]^M$, so this issue does not affect the asymptotics.

Note that in \cite{Viterbi} a somewhat similar approach was taken to
tiling degree 2 de Bruijn graphs ${\cal B}_2^N$.  In the language of
score-based tiles used here, the tilings described in that paper can
be defined by assigning a score of $\phi_i = 0$ when $d_id_{i +1} =
01$ and $\phi_i = 1$ otherwise, with ties broken by choosing $k$ to be
the largest $i$ with $\phi_i = 0$.  As described in \cite{Viterbi},
this gives a tile for $M = 5$ with $E_T = 32$ internal edges (compared
to 42 from the table above for the system defined here).  The tiles
described in \cite{Viterbi} also do not have good asymptotic behavior,
although as mentioned in that paper a generalization of their approach
to match longer sequences can improve behavior in a fixed range of $N$.

\subsection{Asymptotics of  tiles for local pattern-based scores}

We now analyze the asymptotic behavior of the number of broken edges
in the $K$-ary expansion score-based tiles for fixed $K$ as $M
\rightarrow \infty$.  We find that these tiles have the same
asymptotic form for the number of broken edges as the lower bound
(\ref{eq:lower-bound}), multiplied by an overall factor of $c = 2$.
This asymptotic behavior can be understood from a simple idealized
model for the score-based tiles.
We give a simple proof of the asymptotics of the number of broken
edges in this idealized model, and then describe how the tiles just
defined deviate from the idealized model and the consequences of
these deviations for the asymptotic form of the number of broken
edges.  In Section \ref{sec:numerical-asymptotics} we numerically
analyze the score-based tiles and compare to the theoretical asymptotics.

\subsubsection{Idealized model}

Consider an infinite random walk on the de Bruijn graph ${\cal B}_K^M$
underlying a given tile.  Such a random walk will on average traverse
each edge an equal number of times, so the fraction of edges traversed
which are broken will be equal to the fraction of broken edges on the
tile.  In an idealized model of the score-based tile system described
in the previous subsection, we assume that each node of size $M$ has
scores $\phi_i$ which are uniformly and independently distributed
random numbers between 0 and 1.  We assume that the scores are
completely local, so that a node with the score sequence $\phi_1
\cdots \phi_M$ has children with score sequences $\phi_2 \cdots \phi_M
\phi_{M + 1}$ for various values of $\phi_{M + 1}$.  In the idealized
model we can associate the random walk on the de Bruijn graph
underlying the tile with an infinite sequence of numbers $x_1 x_2
\cdots$, each chosen from a uniform random distribution on the set
$[0, 1)$.  The values of $x_i$ correspond to the sequence of scores
$1-\phi_i$ on the nodes encountered in the random walk (while for
  $\phi_i$ we used the convention that the smallest $\phi$ determined
  the level of a node, we reverse this convention for the $x$'s to
  simplify the computations below).  In this
idealization we ignore correlations between the values of $\phi_i$ at
different points $i$, so each $x_i$ is chosen independently from the
uniform distribution.  We associate each ``node'' in the random walk
with a subsequence of $M$ numbers $S_i =$ $x_{i + 1}x_{i + 2} \cdots
x_{i + M}$.  In this model, there are no ties.  An edge ($x_1 \cdots
x_M$, $x_2 \cdots x_M x_{M + 1}$) is broken only if either $x_1$ or
$x_{M + 1}$ is greater than all of $x_2, \ldots, x_M$.  Heuristically,
the chances of either occurrence are $1/M$, giving a probability of
broken edges of $2/M$.

To make this computation more precise, for each consecutive
subsequence of $M$ numbers $S_i =$ $x_{i + 1}x_{i + 2} \cdots x_{i +
M}$ representing a node we assign a level $k$ where $x_{i + k}$ is the
largest value in the subsequence.  We thus have a sequence of values
$k (i)$ defining the positions of the largest $x$'s in each
subsequence $S_i$.  If a value $x_{i + k}$ is the largest $x$ in
subsequence $S_i$, and also is the largest $x$ in $S_{i + 1}$, then $k
(i + 1) = k (i) -1$.  We say that the sequence has a ``broken edge''
at position $i$ when $k (i + 1) \neq k (i) -1$.  We can now show
\vspace*{0.1in}

\noindent
{\bf Theorem 4} In this idealized model, as $M \rightarrow \infty$ the
probability of a broken edge asymptotically approaches $2/M$.

\vspace*{0.1in}

\noindent
{\bf Proof}

Consider the set of ``local maximum'' $x_i$'s which are maximum values
in some subsequence $S_j$ in which they are contained.  There is a
broken edge precisely when the maximum $x$ in one subsequence $S_i$ is
replaced by a new maximum $x'$ in the subsequence $S_{i + 1}$.  Thus,
the frequency of broken edges is the same as the frequency of local
maxima.  For a given number $x$ we can determine the probability that
it is the largest $x$ in a window of size precisely $m$.  For
example, $x_i = x$ is the largest in a window of size precisely 1
when $x_{i -1} > x, x_{i + 1}  > x$.  This occurs with probability
$(1-x)^2$.  $x$ is the largest in a window of size precisely 2 when either
$x_{i-2} > x, x_{i-1} < x, x_{i + 1} > x$ or the symmetric condition
with $x > x_{i + 1}$ but $x_{i -1}, x_{i + 2} > x$.  The probability
that $x$ is largest in a window of size 2 is therefore $2x (1-x)^2$.
A similar set of conditions give $m$ independent cases with
probability $x^{m -1} (1-x)^2$ where $x$ is the maximum in a window of
size $m$.  Integrating over all possible $x$, the probability that a
given $x$ will be largest in a window of precisely size $m$ is
\begin{eqnarray}
p (m) & = & \int_0^1 m x^{m -1} (1-x)^2
= m \int_0^1 x^{m- 1} -2x^m + x^{m + 1}\\
 & = & m \left[ \frac{1}{m}  - \frac{2}{m + 1}  + \frac{1}{m +2}  \right]
= \frac{2}{(m + 1) (m + 2)}  \,.
\end{eqnarray}
the frequency of local maxima is then 
\begin{equation}
f = \sum_{m = M}^{\infty}  p (m) = \frac{2}{M + 1} 
\end{equation}
This proves that in the idealized model the frequency of broken edges
is $2/(M + 1) \approx 2/M +{\cal O} (1/M^2)$.

\vspace*{0.1in}

This result in the idealized model suggests that the smallest fraction
of broken edges we can achieve in a tile based on local patterns in
the node address is asymptotically
$2/M$.  
Now let us consider the relevant differences between the idealized
model and the $K$-ary score-based tiles defined in the previous
subsection.

\subsubsection{Deviations from idealized model}

There are  three primary ways in which the score-based tiles we have
defined deviate from the idealized model.  First, there are some
situations in which different positions have the same score $\phi_i =
\phi_j$ for a fixed node.  Second, the scores are not completely
independent.  Third, the scores $\phi_i$ depend on multiple local
symbols $d_k$, and therefore in particular the score associated with a
point $i$ in a node does not necessarily stay invariant in the
children of that node--- we  have $\phi_{i -1} (v) \neq
\phi_{i} (u)$ along an edge $(u, v)$ whenever the final symbol in $v$
is not $K -1$.

Let us treat these deviations from the idealized model in order.
First, consider the situation of ties.  Generally, ties will increase
the frequency of broken edges.  For example, consider the variation on
the above model where $x_i$ is chosen randomly from a uniform
distribution on the discrete set $\{k/D, 0 \leq k \leq D\}$, and ties
are broken by choosing $k (i)$ to be the smallest $k$ with the maximum
value of $x_{i + k}$ in the set $S_i$.  Then when $M\gg D$, the
frequency of broken edges will be $1/D\gg 1/M$.  (The approach taken
in \cite{Viterbi} is an extreme example of this; their rule for
constructing tiles can be formulated in the language of scores as
described above by taking a score which is either $\phi_i = 0$ if $d_i
d_{i + 1} = 01$ or $\phi_i = 1$ otherwise.  This gives a frequency of
broken edges of 1/4 for large $M$.)

As noted above,  for the $K$-ary score based tiles defined in 
\ref{sec:score-tiles},
ties
only occur for nodes with $d_1, \ldots, d_{n -1} < K -1, d_n = \cdots
d_M = K -1$.  The total number of nodes where such ties occur is
\begin{equation}
N_{\rm ties} = 1 + (K-1)^1 + (K -1)^2 + \cdots (K -1)^{M -1}
=\frac{(K -1)^M -1}{K-2}.
\end{equation}
The fraction of nodes where ties happen therefore decreases
exponentially in $M$ as $[(K-1)/K]^M\ll 1/M$ when $M \rightarrow \infty$.
The consequences of ties are therefore negligible in the asymptotic
behavior of the number of broken edges.

Next, let us consider the other two issues: the lack of independence
between nearby scores and the non-invariance of scores when edges are
followed.  Each of these effects arises from dependence of the score
$\phi_i$ on all the symbols $d_{i + k}$ for any $k > 0$.  From the
form of (\ref{eq:phi-score}) we see that the effect of $d_{i + k}$ on
the score $\phi_i$ is suppressed exponentially as $1/K^k$.  For a node
in a tile of size $M$, the average spacing between the scores of the
$M$ different positions $i$ is $1/M$.  Thus, the impact of $d_{i + k}$
becomes negligible compared to this spacing when $k\gg\log_K M$.
Another way to see this is to note that the smallest $\phi$ will arise
from the longest sequence of 0's in the node which is preceded by the
symbol $K -1$.  As $M \rightarrow \infty$, the average number of
sequences of $n$ zeros in a random node address goes as $M/K^n$.
Thus, we expect ${\cal O} (1)$ sequences of $\log_K M$ 0's but 
${\cal O} (1/M^n)$ sequences of $n + 1$ times this many 0's.  So again, the
number of relevant digits in determining the minimum $\phi$ is of order
${\cal O} (\log_K M)$.  The relevant correlation distance between $\phi$'s is
thus also $\log_K M$.  Similarly, when a new digit sequence giving the
lowest $\phi$ is being shifted in from the right, asymptotically of
order ${\cal O} (\log_K M)$ digits must be shifted in to realize the smaller
value of $\phi$.

The upshot of this analysis is that there are corrections to the
asymptotic form of the idealized model of order $(\log_K M)/M$ when we
consider the score-based tiles described above.  The most important of
these effects is the delay by $\log_K M$ digits in shifting in a new
score $\phi_i$.  When $M-i <\log_K M$, $\phi_i$ will be higher than
the appropriately shifted value in any of the descendants of a given
node due to the boundary condition $d_n = K -1$ for $n > M$ described
below (\ref{eq:phi-score}).  This means that the effective size of the
window associated with the tile is really $M-{\cal O} (\log_K M)$,
which contributes a correction term of order $(\log_K M)/M$ to the
asymptotic form of the fraction of broken edges $2/M$ for the
idealized model demonstrated in Theorem 4.

Because all of these effects are either exponentially suppressed in
$M$ or suppressed by a factor of $(\log_K M)/M$ relative to the
leading $2/M$, none of these effects will modify the asymptotic form
of the number of broken edges for the score-based tiles described
above.  We now verify this analysis with numerical computation.

\subsubsection{Numerical verification of asymptotics}
\label{sec:numerical-asymptotics}

To verify the asymptotic analysis performed in the previous
subsection, and the validity of the approximations made in going from
the score-based tiling system to the idealized model, we have done a
numerical analysis of the number of broken edges for tiles with small
$K$ and reasonably large $M$.  This data corresponds closely with the
theoretical analysis just performed.

\begin{figure}
\begin{center}
\epsfig{file=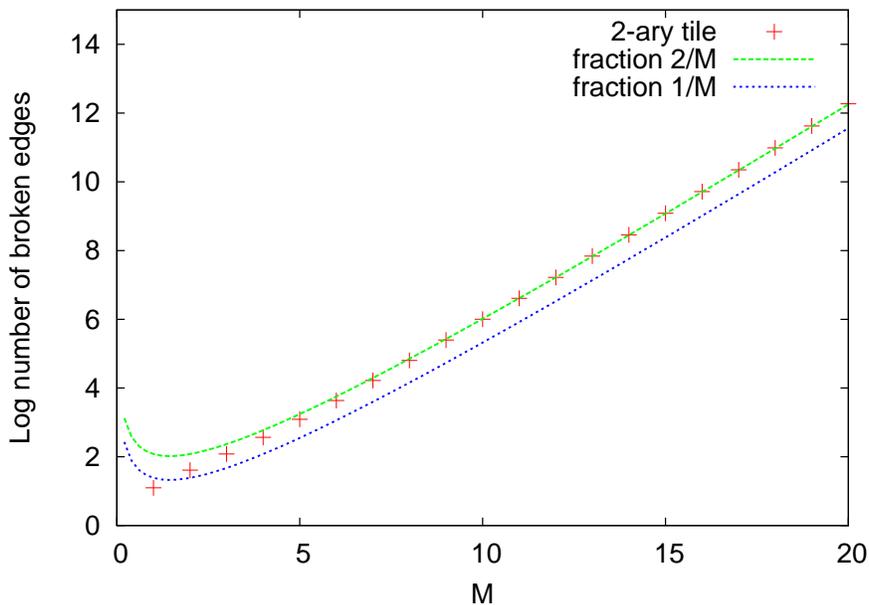,width=12cm}
\end{center}
\caption[x]{\footnotesize (Logarithm of) number of broken edges for
  2-ary expansion score based tile, compared to theoretical asymptotic
  fraction $2/M$ and asymptotic lower bound $1/M$.}
\label{f:numbers-2}
\end{figure}

In Figure~\ref{f:numbers-2} we have graphed the logarithm of the
number of broken edges in the 2-ary expansion score-based tiles for $K
= 2, M \leq 20$.  The number of broken edges in these tiles matches
very closely to the theoretically estimated fraction of $2/M$, and
lies above the asymptotic form of the minimum fraction $1/M$.
Note that for $M > 11$, the number of broken edges exceeds the
fraction $2/M$ of the total number of edges.  This is compatible, however, with the expected 
$(\log_K M)/M$ 
form of the corrections to the asymptotic form computed above.  
For $M > 11$ the computed number of broken edges lies between the
asymptotic form $2/M$ and the log-corrected asymptotic form
$2/(M + 1-(\log_K M)/2)$.

\begin{figure}
\begin{center}
\epsfig{file=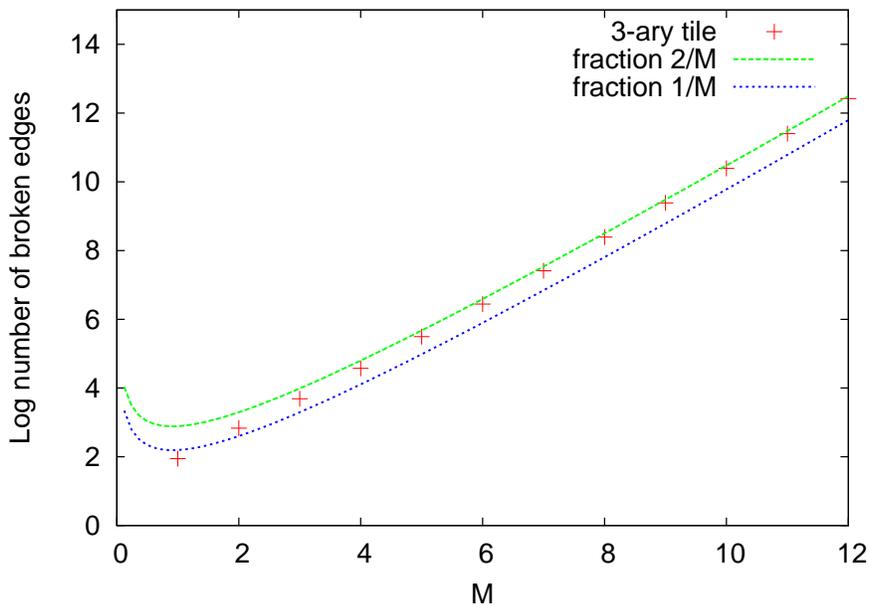,width=12cm}
\end{center}
\caption[x]{\footnotesize (Logarithm of) number of broken edges for
  3-ary expansion score based tile, compared to theoretical asymptotic
  fraction $2/M$ and asymptotic lower bound $1/M$.}
\label{f:numbers-3}
\end{figure}

For tiles with $K = 3$ we have a similar close match to the asymptotic
fraction $2/M$ of broken edges, as depicted in Figure~\ref{f:numbers-3}.

To summarize, from our analysis of an idealized model, we expect that
a fraction of $2/M$ broken edges is asymptotically the optimum which
can be achieved based on a local pattern for level determination.  We
have constructed a family of tiles which realize this asymptotic form,
and are thus asymptotically optimal for local pattern-based tile
structures.  This asymptotic result differs from the lower bound
computed in Section
\ref{sec:asymptotic} by a factor of 2.  We suspect that it is not possible to
find tilings with an asymptotic fraction of less than $2/M$ broken
edges, based on our asymptotic analysis and brute force and greedy
algorithm analysis of small tiles, but we have not proven this
conclusively.  Any mechanism for constructing tiles with an asymptotic
fraction of $c/M$ broken edges with $c < 2$ would require a means of
determining the level of each node based on global properties of the
$dd$ node address, like some kind of global hashing function, rather
than a determination of level based on local patterns such as we have
considered here.

\section{Application to supercomputers}

The results of this paper give not only a theoretical understanding of
how de Bruijn and Kautz graphs can be decomposed into isomorphic
subgraph tiles, but also a concrete approach to constructing such
tilings.  Computer systems containing thousands of individual
processing elements need efficient communication networks to minimize
overhead in passing data between the processors.  Because of their
high degree of connectivity, de Bruijn and Kautz graphs are very well
suited to such large scale processing networks.  The practical problem
of wiring together many processors in such a network is substantially
simplified by the approach of combining multiple processing units into
tiles with isomorphic wiring, and then combining the tiles as we have
described in this paper.

In principle, the methods described in this paper can be used to
design wiring systems for computing systems at a range of scales.  For
any given size of system, the tradeoff between communication and
processing power will affect the choice of degree $K$, and practical
design considerations will affect the choice of tile size $K^M$.  The
results described here should be useful in determining the complexity
of wiring needed for any such system design.  In particular, the lower
bound on edges connecting tiles derived in Section
\ref{sec:asymptotic} gives an absolute lower limit on the complexity
of wiring necessary for such a system.  The explicit tile
constructions given here give concrete examples of wiring patterns
which can be used for such systems.

The feasibility of a practical implementation of the tiling methods
developed in this paper is demonstrated by the Sicortex, Inc.\ family
of cluster computer systems \cite{sc-reference}.  The internal
communication network in these systems is based on a degree $K = 3$
Kautz digraph.  Circuit boards contain processor notes connected in a
stratified subgraph of a degree 3, diameter 3 de Bruijn digraph, so
that a full Kautz graph of any desired size can be wired by connecting
identical boards as described in Theorem 3, taking  
advantage of the parallel routing property.

\section{Conclusions}

In this paper we have developed a systematic approach to partitioning
de Bruijn and Kautz graphs into isomorphic subgraph  ``tiles''
connected by a minimal number of additional edges.  These results
utilize the common mathematical structure underlying de Bruijn and
Kautz graphs, and shed light on the structure of these graphs and
their generalizations.  The tilings we have constructed here have
practical application to the construction of massively parallel
computer systems.

We have given necessary and sufficient conditions for
constructing efficient tilings, we have characterized optimal
tiles in terms of de Bruijn graphs of the size of the desired tile,
and we have constructed an asymptotically optimal class of tilings.
We
have not, however, found a general method for explicitly constructing provably
optimal tiles of arbitrary size, nor have we found a general formula
for the number of internal edges achievable by an optimal tile of
arbitrary size.  We leave these open problems for future work.

\newpage

\bibliographystyle{plain}

\begin{thebibliography}{10}


\bibitem{db}
N.\  G.\ de Bruijn, A combinatorial problem,
{\it Nederl.\ Akad.\ Wetensh.\ Proc.\ Ser. } {\bf A}
49 (1946) 758-764.

\bibitem{Kautz}
W.\ H.\ Kautz, Bounds on directed $(d, k)$ graphs,
{\it Theory of cellular logic networks and machines, AFCRL-68-0668
Final report}, (1968) 20-28.

\bibitem{ii1}
M.\ Imase and M.\ Itoh, Design to minimize a diameter on building
block network, {\it IEEE Trans.\ on Computers}, C-30 (1981) 439-443.

\bibitem{ii2}
M.\ Imase and M.\ Itoh, A design for directed graph with minimum
diameter, {\it IEEE Trans.\ on Computers}, C-32 (1983) 782-784.

\bibitem{rpk}
S.\ M.\ Reddy, D.\ K.\ Pradhan and J.\ G.\ Kuhl, Directed graphs with
minimal diameter and maximal connectivity,
{\it School of Engineering Oakland Univ.\ Tech.\ Rep.,} 1980
 
\bibitem{dch} 
D.-Z.\ Du, F.\ Cao and D.\ F.\ Hsu, de Bruijn digraphs, Kautz
digraphs, and their generalizations,
{\it in} D.-Z.\ Du and D.\ F.\ Hsu, eds., ``Combinatorial Network
Theory'', Kluwer Academic (1996) pp.\ 65-105

\bibitem{bp} 
J.-C.\ Bermond \& C.\  Peyrat, De Bruijn and Kautz networks: A
competitor for the hypercube?, {\it in} F.\ Andr\'e, J.\ P.\ Verjus,
eds., ``Hypercube and Distributed Computers'', North Holland,  (1989) 279-293.

\bibitem{pz}
J.\ Plesnik and S.\ Znam, Strongly geodetic directed graphs, {\it in}
Recent Advances in Graph Theory, Proc.\ Symp., Prague, Academia
  Prague (1975) 455-465.

\bibitem{Viterbi} O.\ Collins, F.\ Pollara, S.\ Dolinar and J.\
Statman, Wiring Viterbi Decoders (Splitting de Bruijn Graphs), 
{\it TDA
Progress Report 42-95}, Jet. Propulsion Laboratory, Pasadena,
California, 1988.

\bibitem{rttv}
J.\ Rolim, P.\ Tvrd\'ik, J.\ Trdlicka and I.\ Vrt'o, bisecting de
Bruijn and Kautz graphs, 
{\it Discrete Applied Mathematics}, 
87-97, 1998.

\bibitem{Leighton}
F.\ T.\ Leighton, {\it Introduction to Parallel Algorithms and
Architectures: Arrays, Trees, Hypercubes}, Morgan Kaufmann, San Mateo
CA (1992).

\bibitem{sc-reference}
N.\  Godiwala, J.\  Leonard, M.\  Reilly, A Network Fabric for  
Scalable Multiprocessor Systems, {\it in} Proceedings of
16th IEEE Symposium on High  
Performance Interconnects, (2008) 137-144.



\end{thebibliography}

\end{document}